\documentclass[%
prl
,preprint
,showpacs
,preprintnumbers
,floatfix
,amssymb, amsmath,nobibnotes, aps
]{revtex4}
\usepackage{bm}%
\usepackage[colorlinks=true,linkcolor=blue]{hyperref}%
\usepackage{epsf}
\expandafter\ifx\csname package@font\endcsname\relax\else
 \expandafter\expandafter
 \expandafter\usepackage
 \expandafter\expandafter
 \expandafter{\csname package@font\endcsname}%
\fi                
\usepackage{graphicx}
\usepackage{dcolumn}
\begin{document}
\title
{
Evidence for particle-hole excitations in the triaxial strongly-deformed
well of $^{163}$Tm 
}
\author
{
N.S.~Pattabiraman$^{1,2}$, Y.~Gu$^{1}$, S.~Frauendorf$^{1}$, U.~Garg$^{1}$,
T.~Li$^{1}$, B.K.~Nayak$^{1}$, X.~Wang$^{1}$, S.~Zhu$^{1,3}$, S.S.~Ghugre$^{2}$,
R.V.F.~Janssens$^{3}$, R.S.~Chakrawarthy$^{4}$, M.~Whitehead$^{4}$,
and A.O.~Macchiavelli$^{5}$ \\\
}

\affiliation
{ 
$^{1}$ Physics Department, University of Notre Dame, Notre Dame, IN 46556, USA \\
$^{2}$ UGC-DAE Consortium for Scientific Research, Kolkata Center,
Kolkata 700 098, India \\
$^{3}$ Physics Division, Argonne National Laboratory, Argonne, IL 60439, USA \\
$^{4}$ Schuster Laboratory, University of Manchester, Manchester M13 9PL, UK \\
$^{5}$ Nuclear Science Division, Lawrence Berkeley National Laboratory,
Berkeley, CA 94720, USA}
\date{\today}
\begin{abstract}
Two interacting, strongly-deformed triaxial (TSD) bands have been identified in
the
$Z = 69$ nucleus $^{163}$Tm. This is the first time that interacting TSD bands
have
been observed in an element other than the $Z = 71$ Lu nuclei, where wobbling
bands have been previously identified. The observed TSD bands in $^{163}$Tm
appear to be associated with particle-hole excitations, rather than wobbling.
Tilted-Axis Cranking (TAC)
calculations reproduce all experimental observables of these bands reasonably
well and also 
provide an explanation for the presence of wobbling bands in the Lu
nuclei, and their absence in the Tm isotopes.
\end{abstract}
\pacs{21.10.Re; 21.60.Ev; 23.20.Lv; 27.60.+j}

\maketitle

Stable asymmetric shapes have been a longstanding prediction of nuclear
structure theory \cite{bohr}. However, experimental evidence for such
{\em triaxial} nuclei has proven difficult to establish. Still, triaxial
shapes have been invoked to interpret a number of experimentally-observed
nuclear structure phenomena such as signature inversion \cite{what1} and
anomalous signature splittings \cite{what2},
chiral twin bands \cite{frau1}, and, most recently, the wobbling
mode \cite{ham}.
Indeed, it has been generally agreed that the most convincing experimental
evidence for stable triaxial shapes is provided by the wobbling mode, recently
established in a number of odd-A Lutetium ($Z = 71$) nuclei
\cite{wob1,wob2,wob3,wob4,wob5,wob6,wob8}. 
The nuclear wobbling motion, akin to
the motion of an asymmetric top,
is indicative of the three-dimensional nature
of collective nuclear rotation \cite{bohr}. In the quantum picture, the
low-spin
spectrum of such a system corresponds to that of the well-known Davydov
asymmetric rotor.  However, the low spin data do not allow a
clear distinction between a rigid rotor and a system that is soft with
respect to triaxial deformation. At high spins, the
sequence of levels that can be associated with the excitation of wobbling
phonons can be better distinguished from soft $\gamma$ vibrations.
The mode is evidenced in the Lu nuclei by families
of strongly deformed (SD) triaxial rotational bands connected to one another
and representing different wobbling phonon quantum numbers $n_{\rm w}$;
bands up to $n_{\rm w} = 2$ have been observed thus far \cite{wob3}.
However, it has been surprising
(and, indeed, somewhat frustrating) that in no other element has this mode
been
observed so far. Indeed, even though a number of SD bands have been reported
in several nearby nuclei (up to 8 in case of $^{174}$Hf!), many of which may
be grouped into possible families based on similarities of their dynamic
moments of inertia, there has been no evidence for connecting transitions
between these bands \cite{hart1,hart2}. Such connecting transitions are 
a {\em sine qua non} condition of wobbling bands and strong
$\Delta J = 1 (0)$, $E2$
linking transitions between the $n_{\rm w}+ 1$ ($n_{\rm w} + 2$) and
$n_{\rm w}$wobbling partners are expected to occur over a large spin range.

We report the observation of two SD bands in the $Z = 69$ nucleus $^{163}$Tm, 
an isobar of $^{163}$Lu, the nucleus with the most extensive experimental
evidence for wobbling bands \cite{wob3}.
We have identified several transitions connecting
the two bands; however, these are unlike the characteristic
transitions between wobbler bands and, instead, are akin to a ``particle-hole
excitation''. Still, this is the first time that two triaxial SD bands with 
interconnecting transitions have been observed in any element other than Lu.
The properties of these bands are well reproduced by calculations in the
framework of the Tilted-Axis Cranking (TAC) model.
Moreover, the calculations provide an explanation of why one observes 
particl-hole eexcitations in the Tm nuclei, but wobbling in the Lu isotopes.

High spin states in $^{163}$Tm were populated via the
$^{130}$Te($^{37}$Cl,4n)$^{163}$Tm reaction, at a bombarding energy of 170 MeV.
The beam was provided by the 88-inch cyclotron facility at the 
Lawrence Berkeley National Laboratory. 
A self-supporting, isotopically enriched target-foil of about
0.5 mg/cm$^2$-thickness was used. To prevent 
contamination and degradation of the target, it was coated with an Aluminum
layer, about 0.04 mg/cm$^2$-thick, on both sides. 
Quintuple- and higher-fold coincidence events were 
recorded with the Gammasphere array~\cite{gs};
at the time of the experiment, the array had 98 active 
Compton-suppressed HPGe detectors. 
A total of about one billion events was accumulated 
and stored onto magnetic tapes for further analysis. 
The data-analysis procedures for developing the level schemes from Gammasphere 
data, and for assignments of spins and parities based on DCO ratio
measurements,
are more-or-less standard by now and only the most pertinent details
are provided here. The data were sorted into three-dimensional and 
four-dimensional histograms~\cite{iucsort,radware} 
and analyzed by projecting double- and 
triple-gated coincidence spectra. The analysis has resulted in 
extensive development of the level scheme of $^{163}$Tm; 
a partial level scheme, relevant to the subject matter of this Letter,
\begin{figure}[htb]
\includegraphics[angle=270, width=12cm]{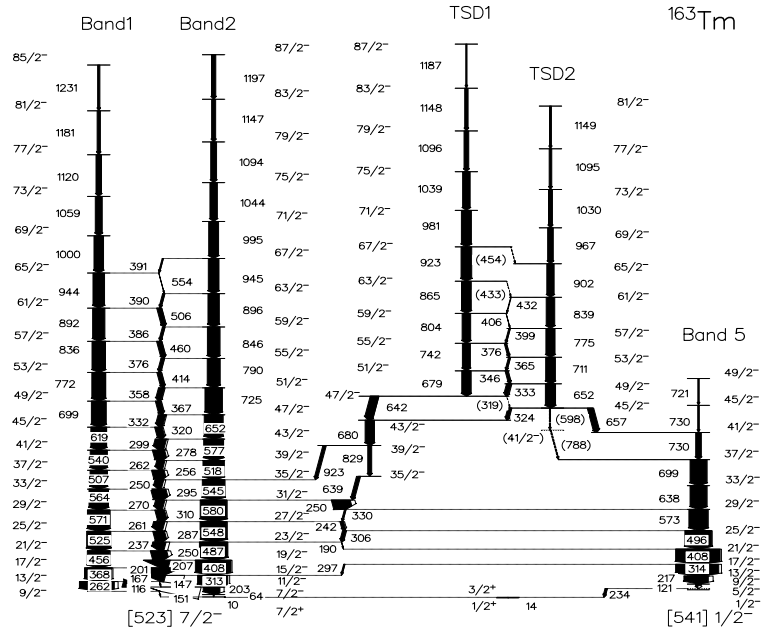}
\caption{\label{scheme}
Partial level scheme of $^{163}$Tm, showing the TSD bands, their interaction, 
and feeding to normal deformed bands. The transition intensities 
are proportional to the thickness of the arrows.
}
\end{figure}
is presented in Fig.~\ref{scheme}.
Supporting coincidence spectra are illustrated in 
Fig.~\ref{spectrum}: the top and middle panels show, respectively, the
$\gamma$-ray transitions in the sequences labeled TSD1 
and TSD2, with an energy spacing ${\Delta}E \sim$~60~keV; the bottom panel
displays many of the ``connecting'' transitions in coincidence with the bands.
Each of these bands is of about 2-3\% of the total intensity in $^{163}$Tm.  

\begin{figure}[htb]
\includegraphics[angle=0, width=12cm]{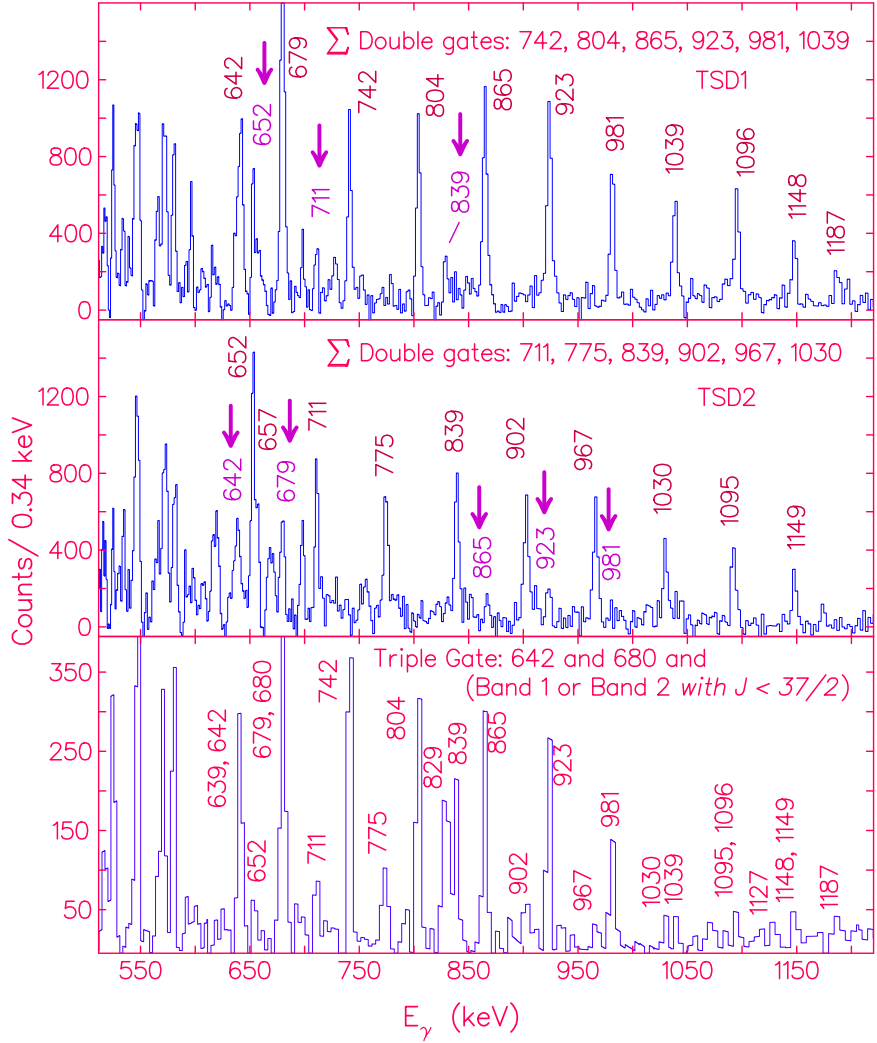}
\vspace*{-1cm}
\caption{\label{spectrum}
Background subtracted coincidence spectra in $^{163}$Tm 
from the summation of spectra with double gates set on 
bands TSD1 (top panel) and TSD2 (middle panel), respectively. The
arrows indicate the appearance of transitions from TSD2 in the TSD1 spectrum,
and {\em vice versa}. The bottom panel provides further evidence for the
``connecting'' transitions.
}
\end{figure}

\begin{figure}[htb]
\includegraphics[angle=0, width=8cm]{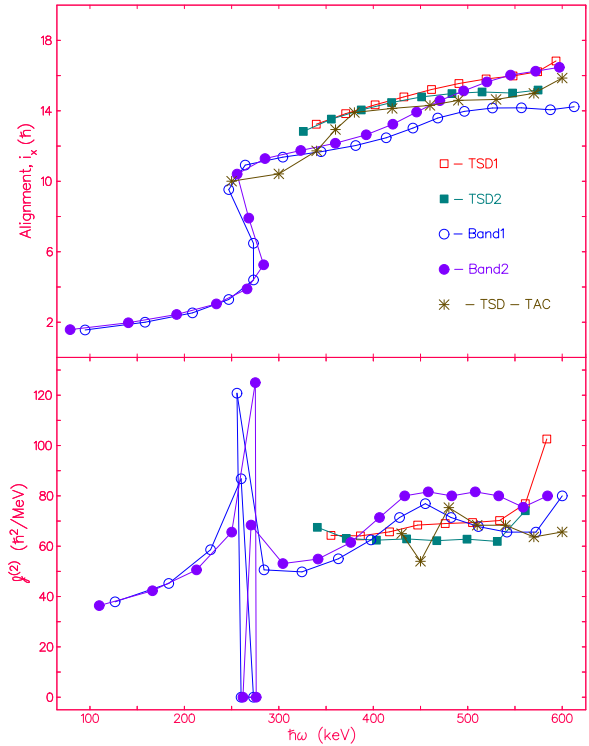}
\vspace{-1cm}
\caption{\label{ixj2}
Alignments $i_x$ (upper panel) and the experimental dynamic moments of inertia 
$J^{(2)}$ (lower panel) for the two TSD bands in $^{163}$Tm 
as a function of rotational frequency.  The reference for 
the alignment is $I_{ref} = {\Im}_0{\omega} + {\Im}_1{\omega}^3$ 
with ${\Im}_0 = 30{\hbar}^2$ MeV$^{-1}$ and 
${\Im}_1 = 40{\hbar}^2$ MeV$^{-3}$. The calculated alignments and the
dynamic moments of inertia for the TSD bands in the TAC model
are also shown.
}
\end{figure}

Angular correlation analyses helped 
in ascertaining the $\Delta J$ = 2 character of the transitions in these bands;
they have all been assigned an $E2$ multipolarity. The linking transitions 
are found to be $\Delta J$ = 1 in character, with 
small possible admixtures, and have been assumed to be of $M1$
multipolarity.
The spin and parity quantum numbers of the two bands are established on the
basis of the multipolarities of the transitions linking them to the
previously-known states in this nucleus \cite{tmold}.
With the established level scheme
and the proposed multipolarity assignments, the alignments, $i_x$,
and the dynamic moment of inertias, $J^{(2)}$,
have been calculated and are plotted  as a function of rotational
frequency in the Fig.~\ref{ixj2}. These plots
\noindent
suggest that the properties of these bands are very similar to those of
other triaxial strongly-deformed (TSD) bands observed in this region and, thus,
have triaxial deformation similarly.
Their SD nature has been
established in a separate DSAM measurement \cite{wang}, where the average
associated transition quadrupole moments were found to be
$Q_{\rm t}$ $\sim$ 8.5 eb in both bands. We note that,
although the dynamic moments of inertia associated with the yrast bands
(labeled Band1 and Band2 in Figs. \ref{scheme} and \ref{ixj2}) are very
similar to those in TSD1 and TSD2, the DSAM measurements \cite{wang} indicate
that their associated $Q_{\rm t}$ moments are significantly smaller~($\sim$~6~eb).

\begin{figure}[htb]
\includegraphics[angle=270, width=12cm]{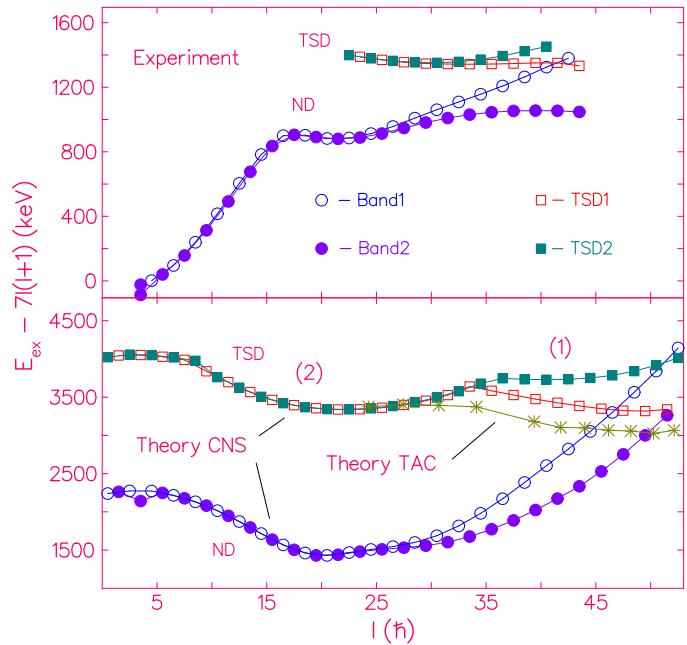}
\caption{\label{excitation}
Excitation energies relative to a rotational reference 
for the two TSD bands and the normal-deformed structures (Band1 and Band2) in
$^{163}$Tm. The top panel shows the experimental energies.
The energies calculated by means of  CNS and
TAC models are presented in the bottom panel. The numbers (1) and (2) with 
the CNS calculation indicate the associated minima from
Fig. 5 (minimum 1 has $\gamma>0^{\circ}$ and minimum 2 has $\gamma<0^{\circ}$).
}
\end{figure}

Fig.~\ref{excitation} shows 
the excitation energies of these TSD bands 
relative to a rigid rotor reference. 
The exhibited pattern is quite different from that observed in the case of
wobbling bands (see, for example, Fig. 14 in Ref. \cite{wob2}), 
and is indicative of the very different nuclear structure associated with
these bands. 
Another major difference from the wobbling bands is that the transitions
between TSD1 and TSD2 are ``interconnecting'', {\em i.e.}, there are
linking transitions
going {\em both} ways between the two bands, whereas for the wobbler bands
the connecting
transitions always proceed only from the band with a higher $n_{\rm w}$ value
to that with a lower $n_{\rm w}$.

To understand the observed properties of these bands and their distinct
differences from the sequences
 in the Lu nuclei, we have performed calculations in the 
frameworks of the configuration-dependent Cranked Nilsson-Strutinsky (CNS)
model \cite{cns} and the Tilted-Axis Cranking model (TAC, 
Shell Correction version SCTAC) \cite{tac}.
The CNS model is a special case of the TAC approach assuming
that the axis of rotation is one of the principal axes. If this axis 
turns out to be stable, CNS calculations provide a solution of the TAC problem.
If it is unstable, one has to use the TAC code to find the tilted solution. 
The technically simpler CNS calculations were carried out first,
using the parameters advocated in Ref. \cite{cns} for the deformed mean field.
The same set was subsequently used for the TAC calculations.
Pairing was assumed to be zero; CNS calculations without pairing have been
successfully applied in numerous cases in the spin range considered
here \cite{ragnar}. 

\begin{figure}[htb]
\epsfxsize 14.0cm
\epsfbox{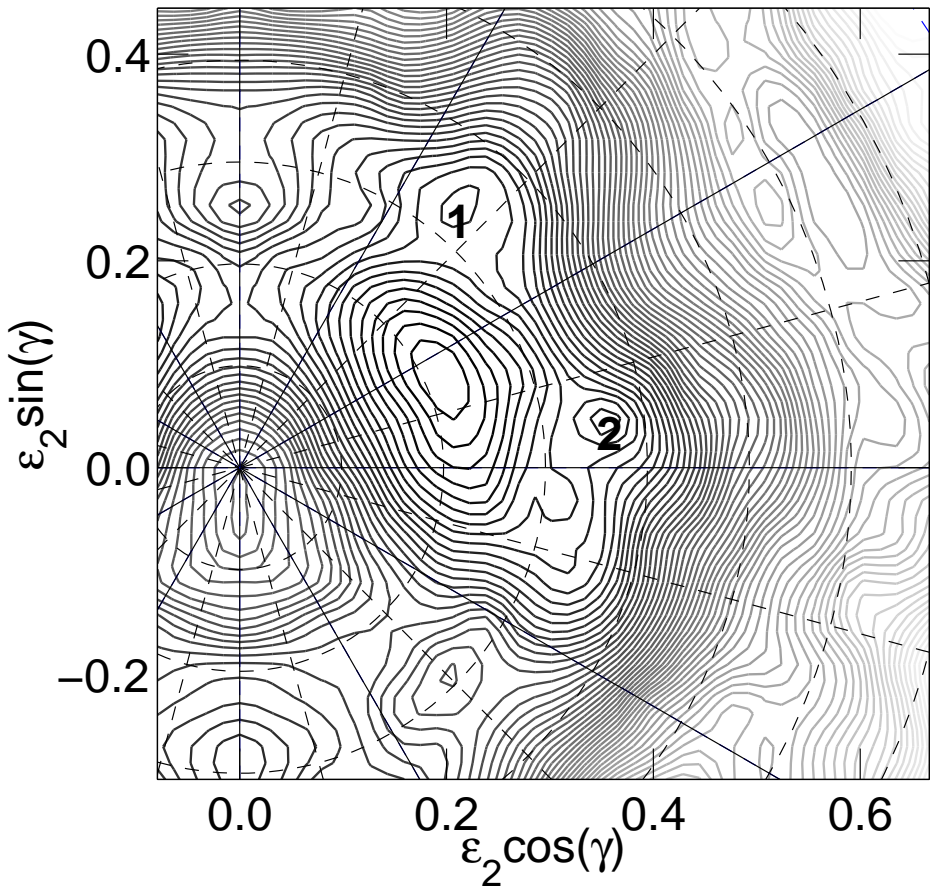}
\vspace{-0.5cm}
\caption{
Potential Energy surface for $^{163}$Tm calculated by means of the CNS
model at $I^{\pi}$ = $63/2^{-}$.
The two TSD minima are marked by  1 and 2. The energy step between the contours
is 0.25 MeV. 
 } \label{PSE}
\end{figure}

Fig. \ref{PSE} presents the CNS energy 
of a configuration with $(\pi,\alpha)=(-,-1/2)$ (see details below) 
as function of the deformation
$\epsilon_2$ and the triaxility parameter $\gamma$
for  $I^\pi=63/2^-$. 
Similar to Ultimate Cranker
calculations in Refs. \cite{wob1,wob2,wob3,wob4,wob5,wob6,wob8,bengt},
the CNS model gives a prolate minimum at normal deformation
($\varepsilon\approx 0.21$), which we refer to as ND, 
and two triaxial strongly deformed  minima, which we refer to 
as TSD. It is worth pointing
 out that, in contrast with previous calculations with $Z> 69$ and
N$\sim$ 94 \cite{wob1,wob2,wob3,wob4,wob5,wob6,wob8},
the $i_{13/2}$ proton level is empty in $^{163}$Tm, 
which means that this level is not essential in
forming the TSD minima. Rather, it is the gap in the neutron spectrum
at $\epsilon_2$ $\approx$ 0.39, $|\gamma|$ $\approx$
17$^{\circ}$ which stabilizes the TSD shape  ({\it cf.} Ref. \cite{bengt}, Fig. 3).

\begin{figure}[htb]
\vspace{1.0cm}
\epsfxsize 12.0cm
\epsfbox{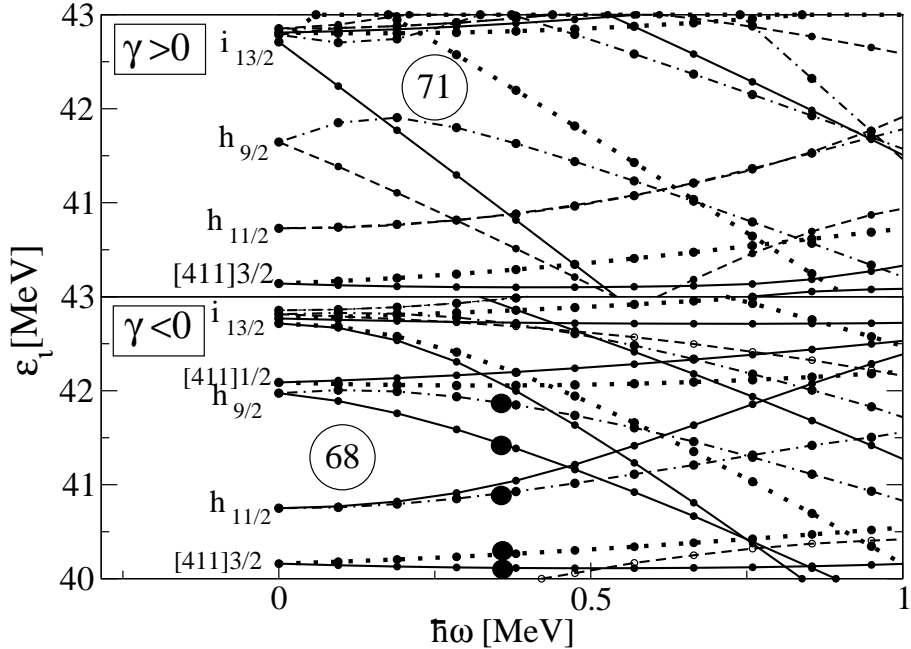}
\caption{ Single-proton routhians as function of
rotational frequency in TSD minima 1(top) and  2 (bottom). The line convention is
$(\pi,\alpha)$= (+,1/2) full, (+,-1/2) dot, (-,1/2) dash, (-,-1/2)
dash dot.
Large filled circles mark occupied levels.     
 } \label{TSDsingle}
\end{figure}

Fig. \ref{TSDsingle} presents the single proton
routhians in both of the TSD minima.
The TSD configuration
that we assign to the observed band TSD1 is indicated by the large filled
circles on the occupied levels. It is the lowest configuration with negative parity
and small signature splitting. There are competing configurations with 
similar energy, which will be discussed below.
The single-proton routhians in the ND minimum (not shown) look 
similar to the ones of Fig. \ref{TSDsingle}, except that the $h_{11/2}$
orbital has a larger splitting between the two signatures, 
as expected for the associated smaller deformation.
We interpret bands 1 and 2 after the backbend as the two signatures of the 
odd proton occupying the $h_{11/2}$ level at the Fermi surface. 
In contrast to the TSD configuration, the proton pair on the $h_{9/2}$
routhians is placed on the $[411]_{1/2}$ routhians  
in the ND configuration.   
Fig. \ref{excitation} compares the
calculated and experimental energies. The observed substantial signature
splitting is consistent with the calculation. 
The measured transition  quadruple moment $Q_t$ for ND bands of $\sim$6 eb
agrees well with the CNS calculation, which gives values around 6.5 eb.        

\begin{figure}[htb]
\epsfxsize 12.0cm
\epsfbox{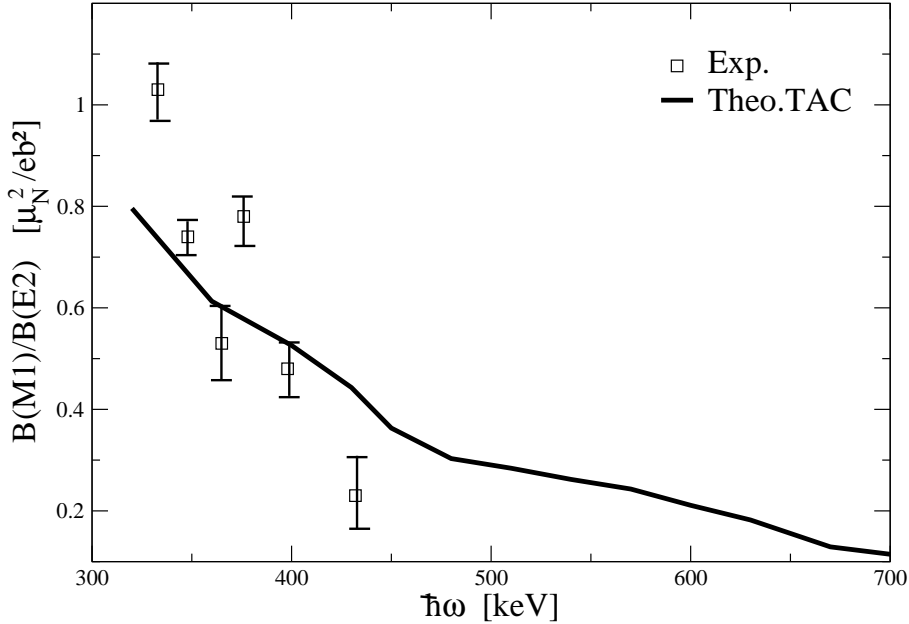}
\caption{B(M1)/B(E2) values as a function of the rotational frequency for the TSD bands.
Squares with error bars
are experimental data; the solid line is the TAC calculation described in
the text.
}
\label{BM1BE2th-exp}
\end{figure}

As seen in Fig. \ref{PSE}, the two TSD minima have nearly the same energy.
It is clear from the bottom panel of Fig. \ref{excitation} that minimum 2 is
energetically favored at low spin and minimum 1 at high spin. The two minima
have almost the same
value of $\epsilon$ and $|\gamma|$, indicating that both are associated
with the same
shape. The axis of rotation is the short one in minimum 1
(with $\gamma>0^{\circ}$), while it is the intermediate one for minimum 2
(with $\gamma<0^{\circ}$). Thus, the CNS calculations suggest
that at $I$ = 24$\hbar$, where minimum 1 goes below minimum 2, the orientation of
the rotational axis flips from the intermediate to the short
  axis. This sudden flip is caused by the inherent assumption in
the CNS model that the rotational axis must be a principal one, and in fact 
indicates that this assumption of rotation about a principal axis is
inappropriate. Therefore, TAC calculations, which do not restrict the
 orientation of the rotational axis, were carried out. As expected,  a tilted
 solution with lower energy was found, 
which smoothly connects  minimum 2 with minimum 1. For $I>23$ the angular momentum
 vector moves away from the intermediate axis toward the short axis. It does not quite
reach it within the considered spin range. (For $I \sim 50$, the angle with the intermediate axis is
still about 20$^\circ$.) This solution is assigned to bands TSD1 and TSD2.
In accordance with the experiment, it
 corresponds to a $\Delta I=1$ band without signature splitting. 
The observed onset of signature splitting at the
 highest spins is consistent with the calculated approach of the TAC
 solution to minimum 1 of the CNS result.
At large frequency, the calculated TSD bands have a lower energy than the ND
ones, which is consistent with the experiment whereby Band 1 crosses TSD1 at
the highest spins.
The TAC calculations for TSD bands give values of the
transition quadruple moment that increase slightly from 8.7 eb at $I=24$ to 
$\sim$ 9.6 eb 
for $34<I<50$, in agreement with the experimental values 
$Q_t \sim$ 8.5 eb. Furthermore, as seen in 
 Fig. \ref{BM1BE2th-exp}, the calculated
 B(M1)/B(E2) ratios of TSD bands agree well with experimental values.
Finally, Fig. \ref{ixj2} demonstrates that the calculations also reproduce the
experimental alignments and the dynamic moments of inertia $J^{(2)}$ very
well.
Thus, all experimental observables for the TSD bands are accounted for
by the TAC calculations.

\vspace{1.5cm}

The configurations that we assign to the bands TSD1 and TSD2 (indicated by the 
large filled circles in Fig. \ref{TSDsingle}) are the lowest with negative
parity and small signature splitting, in agreement with the experiment.
The CNS calculations also predict four other TSD configurations
(termed TSD3, TSD4, TSD5, and TSD6 in the discussion below) at somewhat lower
energy than TSD1 and TSD2. 
The positive-parity configurations TSD3 and TSD4 have the odd proton on
one of the 
$[411]_{1/2}$ routhians and have both $h_{11/2}$ signatures occupied; they
are predicted by the CNS calculations to lie  about 500 keV below TSD1 at spin
20 and have a larger energy above spin 50. However, as can be seen in
Fig. 10 of Ref. \cite{tmold}, some residual proton pair correlations
in the lower-spin part will
disfavor the  configurations TSD3 and TSD4 with respect to TSD1 and TSD2.     
The configurations TSD5 and TSD6, with both signatures of the $h_{11/2}$
orbital occupied and the odd proton 
on one of the two $h_{9/2}$ routhians, would correspond to two well-separated
$\Delta I=2$ sequences with little resemblance to the experiment.  
The favored signature branch, TSD5, is predicted by the CNS calculations to lie 
about 500 keV below TSD1 at spin 20 and
to have a larger energy than TSD1 above spin 40.
We note here that the location of the $h_{9/2}$ orbital has been a longstanding
open problem in calculations using the modified oscillator potential
({\it cf.} the discussion in Ref. \cite{tmold}). On the other hand,
Ref. \cite{witek} demonstrated that
the Woods-Saxon potential (universal parameters) reproduces the position of
this
orbital well for normal deformations. A calculation using the hybrid version of
TAC \cite{dimitrov}, which is a good approximation of the Woods-Saxon
potential, with $\epsilon_{2}$ = 0.4, $\epsilon_{4}$ =0.04, and
$\gamma$ = 17$^{\circ}$, all choices close to the self-consistent values,
places
TSD1 and TSD2 at about the same energy as TSD4 and TSD5 for the low spins, and
below them for the higher spins. This points to possible problems with
the modified
oscillator potential for the TSD shapes which appear to be absent in the
Woods-Saxon potential.

The single particle routhians in Fig. \ref{TSDsingle}
provide a natural explanation for the presence
of  collective wobbling excitations 
in the Lu isotopes with $Z=71$  and their absence in $^{163}$Tm with $Z=69$.
The TSD configurations of nuclei with $Z>69$ belong  typically to minimum 1 with 
$\gamma > 0^{\circ}$ \cite{bengt}.
For $Z=71$, the Fermi level is the $\alpha=1/2$ routhian of 
$i_{13/2}$  parentage in the frequency range
 250 keV$<\hbar\omega<$450 keV. The lowest TSD band is observed in this frequency
range and has $(+,1/2)$. The lowest particle-hole (p-h) excitation of the same parity
lifts the odd proton on to the other signature, $\alpha=-1/2$, of 
this $i_{13/2}$ level, which lies at a relatively high energy
($\sim$ 1 MeV at $\hbar\omega$ = 0.4 MeV).
This brings  the collective wobbling excitation, which has an excitation
energy of about 0.3 MeV, well below the lowest
p-h excitations.  For $Z=69$, however, the two signatures of 
the h$_{11/2}$ state are quite close together ({\it cf.} Fig. \ref{excitation}).   
Therefore, the wobbling excitation lies above the p-h excitations,
likely too high in excitation energy to be populated with observable strength
in the $(HI,xn)$ reaction employed in the present study.
It is also worth mentioning that the relative energy of the collective
wobbling mode and of the p-h excitations in $^{163}$Lu has been studied
by means of the triaxial particle rotor model, where the p-h excitations
have been called the
``cranking mode'' \cite{wob8}. These are found to
be located well above the one-phonon wobbling excitation.
With the level order suggested in Fig. \ref{TSDsingle}, one expects,
for $Z=69$, a  band structure 
similar to the one seen in $Z=71$ at somewhat higher energy; it is obtained by lifting the
last proton from the $h_{11/2}$ into the $i_{13/2}$ orbital. For $Z=73$, several
TSD bands of both parities with  similar energy are expected.

The possibility to experimentally identify a  wobbling band 
is restricted by the  competition of this collective excitation with the
p-h excitations. If the energy of the p-h excitations is high and the 
energy of the wobbling band is  low, it may become the first excited band
above the yrast line. Such a case appears to be realized in the Lu isotopes.
The opposite occurs in $^{163}$Tm. 
The energy for the p-h excitations between the signature partners of the
$h_{11/2}$ orbital is much smaller than the wobbling
energy. In the experiment, only the first excited band,
which corresponds to a p-h excitation, appears to have received sufficient
intensity for observation,
As seen in Fig. \ref{TSDsingle}, the proton level density 
in the frequency range of 300-500 keV is larger for $Z$ = 72 and 73.
 Moreover, we find that
there is a gap at $N$ = 94 in the neutron diagrams, which prevents the neutron
 p-h excitations to compete with the wobbling mode in the Lu isotopes. 
Around $N$ = 102-104, the density of neutron orbitals is high in the relevant
frequency range. This means that, for these nuclides, many low-lying p-h
excitations are possible, and it would be difficult to disentangle a collective
wobbling structure from these many bands. Moreover, the wobbling mode
is expected to be fractionated among the p-h excitations of the same parity.
This would account for the presence of many strongly deformed bands in these
nuclei \cite{hart1,hart2}, 
none of which  shows the characteristics of a wobbling mode. Based on
this observation, Ref. \cite{hart1} suggested
that these nuclides might be less triaxial than the Lu isotopes.
However, as discussed above, the apparent absence
of a wobbling band does not necessarily imply a near-axial shape; indeed, that
would be in contradiction with our
calculations as well as with earlier ones \cite{bengt}.  

In summary, two interacting strongly-deformed triaxial (TSD)
bands have been observed in the $Z$=69 nucleus $^{163}$Tm.
This is the first observation of interacting TDS bands in an element other
than Lu where wobbling bands have been identified. The
observed TSD bands in $^{163}$Tm appear to correspond to particle-hole
excitations, rather than to wobbling. Tilted Axis Cranking
calculations reproduce all experimental observables
for these bands reasonably well. The calculations also provide an 
explanation for the presence of wobbling bands in the Lu isotopes $(Z=71)$ and
their absence in the nearby Tm, Hf, and Ta isotopes ($Z$ = 69, 72, 73).

The authors express their gratitude to Drs. D. Ward, R.M. Clark, and P. Fallon
for their invaluable assistance with the measurements with Gammasphere.
This work has been supported in part by the U.S. National Science Foundation
(Grants No. PHY04-57120 and INT-0111536), the Department of Science and 
Technology, Government of India (Grant No. DST-NSF/RPO-017/98),
the U.S. Department of Energy, Office of Nuclear Physics,
under contract No. W-31-109-ENG-38 and Grant DE-FG02-95ER40934 
and the U.K. Science and Engineering Research Council.

\end{document}